\documentclass[11pt]{article}
\usepackage[usenames,dvipsnames,svgnames,table]{xcolor} % usenames : 16 basic colors,  dvipsnames : other 68 colors, svgnames : another 150 colors. It is used instead of \usepackage{color}. Can find descriptions for color in xcolor package.

\usepackage{enumitem} % for \begin{description}
\usepackage{rotfloat} % For [H] option in sidewaystable
\usepackage{graphicx}
\usepackage{epsfig}
\usepackage{amssymb, amsmath}
\usepackage{verbatim}
\usepackage{natbib}
\usepackage{authblk}
\usepackage{kotex}
\usepackage{enumitem}
\usepackage{multirow}
\usepackage{booktabs} % Top and bottom rules for table
\usepackage{adjustbox}
\usepackage[colorlinks=true,linkcolor=black,citecolor=black,urlcolor=black]{hyperref}%
\usepackage[colorinlistoftodos, textsize=scriptsize]{todonotes} % For making notes. disable option removes all notes.
\usepackage{subcaption}% for \begin{subtable}
\usepackage{algorithm}
\usepackage{algpseudocode}
\usepackage[titletoc,title]{appendix}

\newcommand{\bea}{\begin{eqnarray*}}
\newcommand{\eea}{\end{eqnarray*}}
\newcommand{\bean}{\begin{eqnarray}}
\newcommand{\eean}{\end{eqnarray}}

\newcommand{\bmu}{\boldsymbol{\mu}}
\newcommand{\bS}{\boldsymbol{\Sigma}}

\newcommand{\bx}{\textbf{x}}
\newcommand{\bX}{\textbf{X}}

\newcommand{\forceindent}{\leavevmode{\parindent=1em\indent}}

\begin{document}

\title{Bayesian Nonparametric Classification for Incomplete Data \\ With a High Missing Rate: an Application to Semiconductor Manufacturing Data}
\author[1]{Sewon Park}
\author[1]{Kyeongwon Lee}
\author[2]{Da-Eun Jeong}
\author[2]{Heung-Kook Ko}
\author[1]{Jaeyong Lee}
\affil[1]{Department of Statistics, Seoul National University}
\affil[2]{Samsung Electronics}
\maketitle
\begin{abstract}

During the semiconductor manufacturing process, predicting the yield of the semiconductor is an important problem. Early detection of defective  product production in the manufacturing process can save  huge production cost. The data generated from the semiconductor manufacturing process have characteristics of highly non-normal distributions, complicated missing patterns and high missing rate,  which complicate the prediction of the yield. We propose Dirichlet process - naive Bayes model (DPNB), a classification method based on the mixtures of Dirichlet process and naive Bayes model. Since the DPNB is based on the mixtures of Dirichlet process and learns the joint distribution of all variables involved, it can handle highly non-normal data and can make predictions for the test dataset with any missing patterns. The DPNB also performs well for high missing rates since  it uses all information of observed components.  Experiments on various real datasets including semiconductor manufacturing data show that the DPNB has better performance than MICE and MissForest in terms of predicting missing values as percentage of missing values increases. 

\end{abstract}

\newpage

%%%%%%%%%%%%%%%%%%%%%%%%%%%%%%%%%%%%%%%%%%%%%
\section{Introduction}\label{sec:intro}

\forceindent Missing data occur widely in engineering problems and scientific research. Especially, handling missing value is often the first problem to consider in analyzing the data from the fields of the manufacturing process, social sciences, and biology. Prediction improvement of the yield of semiconductor manufacturing are  the major issues for having stable and consistent manufacturing processes. For the purpose, a huge amount of data consisting of hundreds of variables or factors and millions of observations in the semiconductor manufacturing process is collected and analyzed. However, about 90 percent of data generated from manufacturing process is not recorded due to data storage limit. Metabolomics data contain typically 10-20 percent missing values, which are caused by biological factors such as metabolites being absent, technical reasons such as the limit of detection (LOD), and measurement error \citep{lee2018ns,playdon2019metabolomics}. Furthermore, social scientists conduct empirical research to test or verify theoretical concepts and hypotheses through survey experiments. Some respondents unfortunately provide no information for the survey item. If the missing data are not dealt with appropriately, researchers may draw a wrong conclusion and waste valuable  time and resources.

\cite{rubin1976inference} and \cite{little2019statistical} established three types of missingness mechanisms: (a) missing completely at random (MCAR), (b) missing at random (MAR), and (c) missing not at random (MNAR). When missingness is unrelated to both observed and unobserved variables, the data are called MCAR. When missingness only depends on  observed data, the data are MAR. MAR is a more plausible assumption than MCAR. When missingness depends on unobserved data or the missing value itself, the data are said to be MNAR. An example of the MNAR is censored missing values caused by the LOD \citep{wei2018missing}. In this paper, we present empirical comparisons between the DPNB and other competitors under all missingness mechanisms.

The simplest approach to missing data is to remove incomplete observations or cases. But this method may lead to biased conclusions and lose some useful information.
Another approach is the imputation which substitutes missing data for plausible values generated from statistical learning. Imputing missing data may be a more reasonable way than discarding missing values but not necessarily give better results. 

We review several popular imputation techniques for estimating missing values. The basic method is the mean imputation, which replaces missing values with the mean of the observed values in a certain variable. The mean imputation is simple to apply but underestimates the variance and produces biased results under both MAR and MNAR \citep{little2019statistical}. Another method is the multivariate imputation by chained equations (MICE) developed by \cite{buuren2010mice}. The MICE as a type of  multiple imputations \citep{rubin1976inference} constructs separate conditional models for each incomplete variable and iteratively imputes the missing values. Gaussian mixture model (GMM) is one of the most widely used  model-based imputation methods \citep{ghahramani1994supervised,lin2006fast,williams2007classification}. Model-based methods assume the joint distribution of all variables in the data and estimate parameters of the distribution \citep{das2018handling}. GMM on incomplete data imputes missing values using conditional distribution properties of a multivariate normal distribution. 

Contrary to the model-based methods, there are imputation strategies based on machine learning algorithms that do not rely upon distributional assumption on the data.  \cite{burgette2010multiple} and \cite{stekhoven2012missforest} designed decision tree-based imputation methods using classification and regression tree  and random forest, respectively. Another non-parametric approach is K-Nearest Neighbors (KNN) based imputation \citep{troyanskaya2001missing}. \cite{caruana2001non} and \cite{bras2007improving} improved the accuracy of estimated missing values using an iterative process. The aforementioned imputation methods from machine learning can deal with both numerical and categorical data as well as mixed-type data. Thus, these methods have been commonly used in various fields related to missing data problems. 

Deep learning-based imputation methods have been proposed in different neural network architectures. \cite{sharpe1995dealing}, \cite{gupta1996estimating}, and \cite{silva2011missing} reconstructed missing values by using feed-forward neural networks (FNN). \cite{bengio1996recurrent}, \cite{che2018recurrent}, and \cite{kim2018temporal} introduced recurrent neural networks to handle incomplete sequential data. \cite{vincent2008extracting} and \cite{gondara2018mida} designed  deep generative models combining denoising autoencoders (DAE) to create the clean output from the noisy input considered as missing data. \cite{yeh2017semantic} and \cite{yoon2018gain} provided modified generative adversarial nets (GAN) for filling missing values or regions.

% Imputation techniques described above focus on estimating missing values and cannot conduct subsequent statistical analyses such as classification or regression. Most classification approaches on incomplete data designed combinations of a predictive model and imputation strategy. In other words, they used to transform missing data to complete cases and build a classifier in the training phase \citep{batista2003analysis, saar2007handling, garcia2010pattern, ghorbani2017performance, bertsimas2017predictive}. However, missing data may exist not only in the training set but also in the test set. In the literature, there are a few solutions for addressing  the test set which has missing values  $\to$ the problems without using the simple imputation techniques

% \bch General frameworks cannot perform classification in the testing process  if all variables in the test set are empty. $\to$ In the testing process, general frameworks can only perform classification by using mean or median imputation methods obtained in the training process if all observations of some variables in the test set are empty. \ech In the literature, there are a few solutions for addressing \bch the test set which has missing values  $\to$ the problems without using the simple imputation techniques \ech \citep{wang2010classification, garcia2010pattern, jiang2020logistic}. \ech \footnote{ 동시에 추론이 가능하다는 식으로 간단한 문장 추가} 

In this paper, we propose a new combined method that performs both imputation and classification tasks. Our proposed method has several advantages over the existing methods. First, the class conditional distribution is an infinite Gaussian mixture model instead of a Gaussian used in a standard naive Bayes classifier. The proposed classifier can construct flexible decision boundaries and learn various types of non-linear decision boundaries. Second, the proposed imputer is more accurate than other imputation methods on incomplete data with high missing rates. In other words, as the percentage of missing values increases, the imputation technique based on class conditional distribution outperforms state-of-the-art methods under both MCAR and MAR assumptions. While other imputation methods utilize partial observed information by missing patterns and algorithms, the DPNB uses all information of observed components. Thus, this gap will be more apparent when the rate of missing data is high. Third, the DPNB is capable of predicting the labels of a set of new cases which have any missing patterns from a predictive model. Imputation techniques described above focus on estimating missing values and cannot conduct subsequent statistical analyses such as classification or regression. Most classification approaches on incomplete data designed combinations of a predictive model and imputation strategy. They used to transform missing data to complete cases and build a classifier in the training phase. However, missing data may exist not only in the training set but also in the test set.
Our method can address the test set which has missing values with any missing patterns. Experiments also show that the proposed method gives better classification accuracy on multiple datasets with high missing rates from the UCI Machine Learning Repository.

The remainder of the paper is structured as follows. Section 2 provides preliminary notions of imputation through finite Gaussian mixture models and the Dirichlet process prior as a key factor in our proposed method. In section 3, we propose an extension of the well-known Mixture Discriminant Analysis (MDA), which uses an infinite Gaussian mixture model on incomplete data.
Section 4 shows empirical results on the accuracy of imputation and classification in different settings. A real data example is presented in Section 5. We end this paper in section 6 with conclusions.

%%%%%%%%%%%%%%%%%%%%%%%%%%%%%%%%%%%%%%%%%%%%%%%%%%%%%%%%%%%%%%%%%%%%%%%%%%%%%%%%%%%%%%%%%%%%%%%%%%%%%%%%%

\section{Preliminaries}\label{sec:pre}

\forceindent In this section, we review the Gaussian mixture models on missing data and the Dirichlet process, which are core concepts of our proposed method.

\subsection{Gaussian mixture models for missing Data}\label{subsec:miss}

\forceindent Algorithms for Gaussian mixture models on missing data have been studied in the last few decades. \cite{ghahramani1994supervised} used the Expectation-Maximization (EM) algorithm to find parameter values and missing components maximizing the likelihood of Gaussian mixture models. \cite{zhang2004bayesian} developed a Bayesian approach of mixture models using Gibbs sampler. This method utilizes full conditional distributions to obtain the joint posterior distribution of parameters and missing values.  \cite{williams2007classification} introduced variational inference based on the mean-field approximation for Bayesian mixture models. Both missing values and parameters are iteratively updated until the evidence lower bound (ELBO) converges.

We focus on the estimation of a Gaussian mixture model, permitting simultaneous inference of missing values through Gibbs sampling is a Markov Chain Monte Carlo (MCMC). Let $\textbf{x}_i, i = 1, \ldots, n$ be $n$ independent $p$-dimensional observations from the mixture distribution consists of $H$ Gaussian components. 
We partition an observation into two components $ \textbf{x}_i =  \{\textbf{x}_i^{o_i}, \textbf{x}_i^{m_i}\}$, where $o_i \subset \{1,2,\ldots, p\}$ is an index set of observed variables and $m_i \subset \{1,2,\ldots, p\}$ is an index set of missing variables . That is, $\textbf{x}_i^{o_i}$  and  $\textbf{x}_i^{m_i}$ indicate the observed components and  missing components from the $i$th observation $\textbf{x}_i$, respectively. We can express the mixture distribution as follows: 
\begin{equation}
    f(\textbf{x}_i) = \sum_{h = 1}^H w_h \mathcal{N}_p (\textbf{x}_i | \bmu_h, \bS_h) = \sum_{h = 1}^H w_h \, \mathcal{N}_p \left( \begin{bmatrix} 
\bx_i^{o_i}  \\
\bx_i^{m_i} 
\end{bmatrix} \Bigg| \begin{bmatrix} 
\bmu_h^{o_i}  \\
\bmu_h^{m_i}
\end{bmatrix} , \begin{bmatrix} 
\Sigma_h^{o_i,o_i}  & \Sigma_h^{o_i,m_i}\\
\Sigma_h^{m_i,o_i}  & \Sigma_h^{m_i, m_i}
\end{bmatrix} \right)
\end{equation}
Here, $w_h$ is the non-negative mixing proportion and sum up to one, i.e., $\sum_{h=1}^H w_h = 1$.
% Let us introduce the cluster assignment, $\textbf{z}_i = (z_{i1}, \ldots, z_{iH})$ such that $z_{ih} = 1$ if an observation $\mathbf{x}_i$ belongs to cluster $h$ satisfying $\sum_{h = 1}^H z_{ih} = 1$. The marginal distribution for $z_{ih}$ is specified in terms of  the non-negative mixing proportion $w_h$, such that $p(z_{ih} = 1 ) = w_h$ satisfying $\sum_{h = 1}^H w_h = 1$. 
% $z_{i}^k = h$ denotes the cluster assignment of observation $\mathbf{x}_i$ in the class $k$ to cluster $h$ and $I(\cdot)$ is an indicator function. 
% A natural conjugate prior distributions over $(\bmu_k, \bS_k)$ is the multivariate normal-inveser wishart distibution:
% \begin{align*}
% \bS_k &\sim \mathcal{IW}( \textbf{B}_0, \nu_0),  \\
% \bmu_k | \bS_k &\sim \mathcal{N}_p(\textbf{m}_k, \bS_k/ \tau_0 ), \end{align*}
% where $\tau_0 > 0$. We put a dirichlet prior over the vector of mixing proportions, $\textbf{w} = (w_1, \ldots, w_k)$
% \begin{equation*}
% \textbf{w}  \sim \mathcal{D}(\boldsymbol{\alpha}= (\alpha_1, \ldots, a_K)). 
% \end{equation*}
% Futehrmore, we introduce auxiliary variables, $\textbf{z}_i = (z_{i1}, \ldots, z_{iK})$ to simplify and improve the sampler. This auxiliary variables imply to which mixture component each observation belongs and  prior distribution over them, $\textbf{z}_i$ is a multinomial distribution with K categories and one trial. 
% \begin{equation*}
%     p(\bx_i | \textbf{z}_i = k) = p(\bx_i | \bmu_k, \bS_k), \quad \textbf{z}_i \sim \mathcal{M}(1, \textbf{w}).
% \end{equation*}

To implement the Gibbs sampling for mixture models with incomplete data, we need the full conditional posterior distributions of model parameters and missing values. In this paper we only cover the full conditional posterior for missing values and skip other parameters of mixture models; see the paper by \cite{franzen2006bayesian} for more details. If the data come from a multivariate normal distribution with mean $\bmu$ and covariance $\bS$, a full conditional density for missing values based on the observed data can derive easily the following equations \eqref{eq:normalcond} by using a special property of multivariate normal distribution.
\begin{equation}
\begin{aligned}
    f(\bx_i^{m_i} | \bx_i^{o_i}, \text{others}) \sim \mathcal{N}_{|m_i|}(\bx_i^{m_i}; \bmu^{m_i|o_i}, \bS^{m_i | o_i}), \\
    \bmu^{m_i|o_i} = \bmu^{m_i} + \bS^{m_i, o_i} (\bS^{o_i, o_i})^{-1}( \bx_i^{o_i} - \bmu^{o_i}), \\
    \bS^{m_i|o_i} = \bS^{m_i, m_i} - \bS^{m_i, o_i}(\bS^{o_i, o_i})^{-1}\bS^{o_i, m_i}. 
\end{aligned}
\label{eq:normalcond}
\end{equation}
Missing values are filled with samples drawn from its full conditional distribution and combined with the fixed observed values to update subsequently other parameters from full conditionals. This updating process is called the multivariate normal imputation, first implemented by \cite{schafer1997analysis}, which is one of the multiple imputation methods. In the case of the mixture models,  we assume all data points are generated from  the mixture  of multivariate normal distributions. A observation with the missing values belong to $h$th mixure component is imputed by sampling full conditional distribution for $\bx_i^{m_i}$ given $\bmu_h$ and $\bS_h$ associated with $h$th mixture component. Mixture components that indicate clusters are determined by latent auxiliary variables. See \cite{zhang2004bayesian}.

\subsection{Dirichlet process}

\forceindent Dirichlet process (DP) introduced by \cite{ferguson1973bayesian} is the most popular Bayesian nonparametric model and used in many applications for clustering in the last two decades. Let $\mathcal{X}$ be a measurable space and $\mathcal{B}$ the Borel $\sigma$-field of subsets of $\mathcal{X}$. Then we say that the random probability measure $G$ on $(\mathcal{X}, \mathcal{B})$ follows a Dirichlet process with a concentration parameter $\alpha > 0$ and a baseline probability measure $G_0$, denoted by $G \sim DP(\alpha, G_0)$, if for every finite disjoint partition $A_1, \ldots A_k$ of $\mathcal{X}$,
$$
(G(A_1),\ldots, G(A_k)) \sim \text{Dir}(\alpha G_0(A_1),\ldots,\alpha G_0(A_k)),
$$
where $\text{Dir}$ denotes the Dirichlet distribution. Dirichlet process can be represented in three different ways: (1) p\'{o}lya urn scheme \citep{blackwell1973ferguson} (2) Chinese restaurant process \citep{aldous1985exchangeability} (3) stick-breaking process \citep{sethuraman1994constructive}. 

\cite{sethuraman1994constructive} also showed that its realizations are discrete almost surely, even if $G_0$ is a  continuous distribution. The discreteness of the DP implies that it is unsuitable prior for data generated from continuous distributions. To eliminate this drawback, \cite{antoniak1974mixtures} adopted Dirichlet
process mixture models (DPMM) having the following hierarchical model formulations: for $i = 1, \ldots,n, $
\begin{equation*}
\begin{aligned}
 \textbf{x}_i | \theta_i &\stackrel{ind}{\sim} f(\textbf{x}_i | \theta_i), \\
     \theta_i &\stackrel{iid}{\sim} G, \\
     G &\sim DP( \alpha,  G_0),
\end{aligned}
\end{equation*}
where $f$ is a parametric density function. The DPMM can be expressed as a limit of finite mixture models, where the number of mixture components is taken to infinity \citep{teh2006hierarchical}. For example,  if a base measure $G_0$ is a multivariate normal Inverse-Wishart conjugate prior and $f(\textbf{x}_i | \theta_i)$ is multivariate normal, the DPMM can be an infinite normal mixture model. The advantage of the DPMM is the number of mixture components is not fixed in advance of fitting the data and automatically infer the
number of clusters from the data unlike finite normal mixture models. Various algorithms have been developed for posterior inference of the DPMM such as marginal Gibbs sampling \citep{maceachern1994estimating, escobar1994estimating,escobar1995bayesian,bush1996semiparametric,neal2000markov}, conditional Gibbs sampling \citep{ishwaran2001gibbs, walker2007sampling, kalli2011slice, ge2015distributed}, split-merge MCMC sampling \citep{jain2004split}, sequential updating and greedy
search (SUGS) algorithms \citep{wang2011fast}
and variational approximation \citep{blei2006variational}.

%%%%%%%%%%%%%%%%%%%%%%%%%%%%%%%%%%%%%%%%%%%%%%%%%%%%%%%%%%%%%%%%%%%%%%%%%%%%%%%%%%%%%%%%%%%%%%%%%%%%%%%%%

\section{Proposed model}\label{sec:model}

% naive Bayes 모형 언급
% DP-naive Bayes 모형 소개

\forceindent In this section we propose a new classification approach for handling incomplete data based on the Dirichlet process mixture model. We call the proposed method as Dirichlet Process-Naive Bayes model (DPNB).

\subsection{Generative model with DPMM}
Generative models employ the Bayes theorem to build a classifier. Let the input or feature vector be $\bX$ and the class labels be $Y$. We need the density of $\bX$ conditioned on the class $k$, $P( \bX = \bx | Y = k)$ and the prior probability, $P(Y = k)$ to compute the posterior probability:
\begin{equation}
    P(Y = k | \textbf{X}  = \bx) = \dfrac{P(Y = k) \cdot P( \bX = \bx | Y = k)}{\sum_{l  = 1}^K P(Y = l) \cdot P( \bX = \bx | Y = l)}.
    \label{eq:bthm}
\end{equation}
Then, input $\bx$ is assigned to the class having the highest posterior probability.  Typical examples of generative models include linear discriminant analysis (LDA), quadratic discriminant, analysis (QDA), and Gaussian naive Bayes (GNB). The three  models assume multivariate normal densities for $P( \bX = \bx | Y = k)$. Another example is to assume that the class-conditional density of $\bX$ is a finite mixture of normals. This method is called mixture discriminant analysis (MDA) \citep{hastie1996discriminant, fraley2002model}.  Instead of using a finite mixture of normals, we propose to use the Dirichlet process mixture model, as the number of clusters is automatically determined and not limited. 

We consider a binary classification problem and assume class labels $Y \in \{0, 1\}$  has a binomial distribution with parameters $n$ and $p$. If we have $K$ $(>2)$ classes, $Y \in \{1,\ldots, K\}$ is assumed to have a multinomial distribution with $n$ and $\textbf{p} = (p_1, \ldots, p_K)$. The distribution of $\bX$ conditioned on the class $k$ is modeled by the following hierarchical formulation for DPMM:
\begin{equation}
\begin{aligned}
\bx_i | y_i = k &\stackrel{ind}{\sim} \mathcal{N}(\bx_i^k; \bmu_i^k, \bS_i^k), \\
(\bmu_i^k, \bS_i^k)  &\stackrel{iid}{\sim} G_k, \quad i = 1,2,\ldots, n_k ,\\
G_k  &\sim DP(\alpha, M_k), \quad k = 0, 1,
\end{aligned}
\label{eq:dpmm}
\end{equation}
where $\bx_i^k$ is the $i$th feature vector and $\boldsymbol \mu^k$ and  $\boldsymbol \Sigma^k$ are parameters of a normal distribution which belongs to only class $k$. Here, the base measure for class $k$, $M_k$ is the conjugate multivariate normal–inverse Wishart distribution, i.e. $M_k := \mathcal{N}(\bmu^k; \mathbf{m}_0^k, \bS^k/\tau_0^k) \cdot \mathcal{I}\mathcal{W}(\bS^k; \mathbf{B}_0^k,\nu_0^k)$. Then the class-conditional density of $\bX$ is given by
\begin{equation}
    P(\bX = \bx | Y = k) = \sum_{h = 1}^{\infty} w_h^k \mathcal{N}(\bx ; \bmu_h^k, \bS_h^k), \quad k \in \{0,1\}
    \label{eq:cond}
\end{equation}
using posterior samples for mixing proportion and cluster specific parameters. We use the improved slice sampler suggested by \cite{ge2015distributed} to generate samples from the posterior distribution. We describe the detailed MCMC algorithm for the DPMM on incomplete data in \autoref{sec:appendA}.  Since $Y$ has a binomial distribution, marginal probabilities over classes are estimated by
\begin{equation}
 P( Y = k) = \frac{n_k}{n}, \quad k \in \{0,1\},
 \label{eq:mag}
\end{equation}
where $n_k$ denotes the number of observations belonging to the class $k$.
Putting \eqref{eq:cond} and \eqref{eq:mag} together in the equation \eqref{eq:bthm}, we can produce the following posterior probabilities for classes: 
\begin{equation*}
    P(Y = k | \bX = \bx) = \frac{ n_k \times \sum_{h = 1}^{\infty}w_h \mathcal{N}(\bx ;  \bmu_h^k, \bS_h^k)}{\sum_{l \in \{0,1\}} \left( n_l \times \sum_{h = 1}^{\infty} w_h \mathcal{N}(\bx ; \bmu_h^l, \bS_h^l) \right)}.
\end{equation*}
We compute posterior probabilities for all $k$ and  choose only one class associated with the highest probability. 

\begin{figure}[ht!]
  \centering

  \begin{subfigure}{0.94\linewidth}
    \centering
    \includegraphics[width=\linewidth]{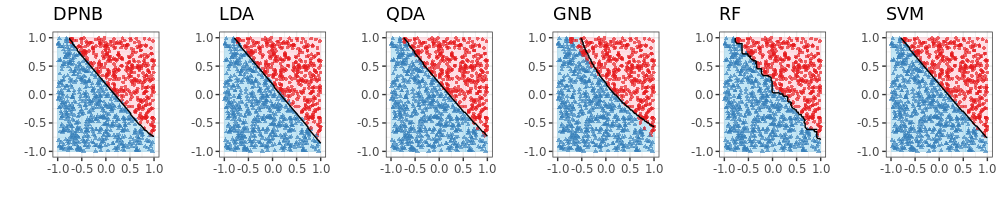}
    \caption{Linearly separable dataset}
  \end{subfigure}

  \begin{subfigure}{0.94\linewidth}
    \centering
    \includegraphics[width=\linewidth]{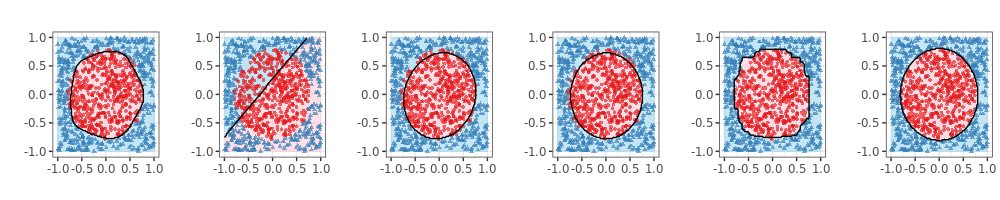}
    \caption{Circle dataset}
  \end{subfigure}  
  
  \begin{subfigure}{0.94\linewidth}
    \centering
    \includegraphics[width=\linewidth]{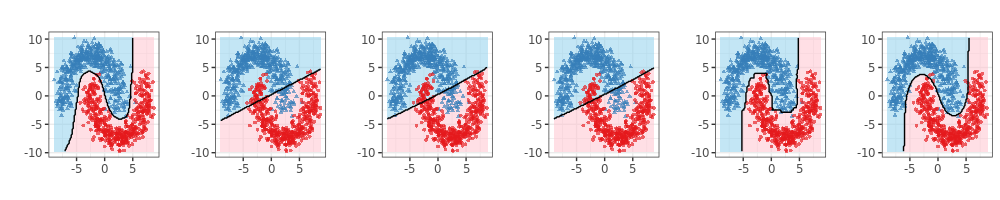}
    \caption{Two moons dataset}
  \end{subfigure}  
  
  \begin{subfigure}{0.94\linewidth}
    \centering
    \includegraphics[width=\linewidth]{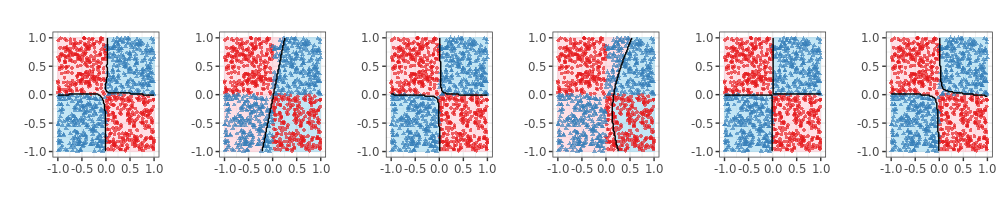}
    \caption{XOR dataset}
  \end{subfigure}  
  
  \begin{subfigure}{0.94\linewidth}
    \centering
    \includegraphics[width=\linewidth]{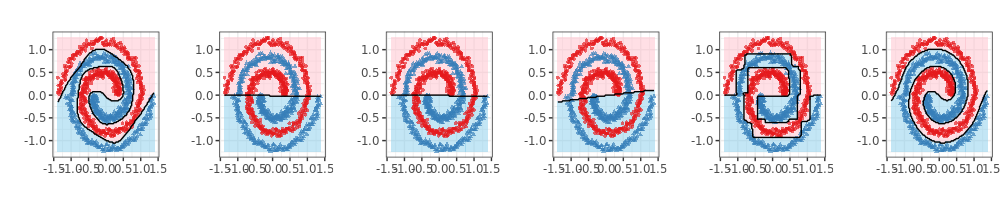}
    \caption{Two spirals dataset}
  \end{subfigure}  
  
  \caption{Comparison of decision boundaries of the DPNB model using the approximate prediction rule and five classifiers.}  
  \label{fig:db}
\end{figure}

% classifier 특징
Both LDA and QDA assume multivariate normal densities for $P( \bX = \bx | Y = k)$ and GNB assumes that the features, $\bx_1, \bx_2, \ldots, \bx_p$ are conditional independent on $Y$ and $P(\bx_j | Y = k)$ is univariate normal density:  
$$
P( \bX = \bx | Y = k) = \prod_{j = 1}^p P(\bx_j | Y = k).
$$
Since three generative models are designed to different covariance structures of multivariate normal distributions, they have their respective classifiers.
The decision boundaries for LDA are linear, while those for QDA are quadratic. GNB has not necessarily linear classification according to the data. However, MDA allows close approximation of not only linear but also nonlinear decision boundaries since mixture models can approximate arbitrary continuous distributions \citep{fraley2002model,wang2010classification}. \autoref{fig:db}
shows that DPNB more accurately approximates most decision boundaries than do LDA, QDA, and GNB. In other words, DPNB is a much more flexible classifier than them. Furthermore, the classifier of DPNB provides comparable results with that of Support Vector Machine (SVM) and Random Forest (RF).

\subsection{Imputation and prediction strategy}
\forceindent The DPNB can  make inferences simultaneously on both unknown quantities and missing values. The same imputation approach described in the previous subsection \ref{subsec:miss}  is capable of being applied in the mixture models with infinite components. For estimating \eqref{eq:cond}, DPNB should divide the data into subsets where each subset belongs to only one class according to the inference process. Then it substitutes separately missing values with plausible values generated from samplers \eqref{eq:normalcond} based on respective subsets. We expect that imputation based on subsets is more accurate than that based on the full data. Because it is easy to find easily probable values estimated from homogeneous input data which belongs to only one class. If we spotlight imputation problems  in the classification data, complete subsets are put together into a complete full dataset. In practice, the empirical results on 4 UCI datasets provide that this split and merge imputer has better performance than state-of-the-art imputation algorithms as the missing rate increases.

In reality, missing data may occur both training and test dataset. DPNB can make predictions for classes even if all elements of some features in the test set are absent. Let a new observed input vector denote  $\bx_\star^{o_{\star}}$
and  a new predicted class label $Y^{\star}$.  The prediction rule for DPNB is given by
\begin{equation}
P(Y^{\star} = k | \bx_\star^{o_{\star}}) = \dfrac{P(Y^{\star} = k) \cdot P( \bx_\star^{o_{\star}} | Y^{\star} = k)}{\sum_{l \in \{0, 1\}} P(Y^{\star} = l) \cdot P( \bx_\star^{o_{\star}} | Y^{\star} = l)}, \quad k \in \{0, 1\}.
\label{eq:missbayes}
\end{equation}
We need to compute the predictive density of $\bx_\star^{o_{\star}}$ conditioned on the class $k$ to complete the equation \eqref{eq:missbayes}, the posterior probability of the new class label. It can be approximated by
using the Monte Carlo Integration  from the posterior samples in the training process as shown in \eqref{eq:predn}. Therefore, DPNB can build classifiers regardless of missingness and missing rate on both training set and test set. Experiments also support that it predicts classes more accurately than do competitive models. 
\begin{equation}
\begin{aligned}
    P( \bx_\star^{o_{\star}} | Y^{\star} = k) &= \int P( \bx_\star^{o_{\star}}, \bx_\star^{m_{\star}} \,| Y^{\star} = k)\,\, d\bx_\star^{m_{\star}}  \\
    & = \int P( \bx_\star^{o_{\star}}, \bx_\star^{m_{\star}} \,|\, \bmu, \bS, Y^{\star} = k)  \pi(\bmu, \bS \, | Y^{\star} = k ) \,d \bmu \,d \bS d\bx_\star^{m_{\star}} \\
    % & = \int  \sum_h  P( \bx_\star^{o_{\star}}, \bx_\star^{m_{\star}} \, | \bmu_h, \bS_h,   Y^{\star} = k) \pi(\bmu_h, \bS_h \, | Y^{\star} = k ) \pi(z = h \, | Y^{\star} = k) \,d \bmu_h \,d \bS_h \, d\bx_\star^{m_{\star}}  \\
    & = \int  \sum_h \pi_h^k  \, P( \bx_\star^{o_{\star}}, \bx_\star^{m_{\star}} \, | \bmu_h, \bS_h,   Y^{\star} = k) \pi(\bmu_h, \bS_h \, | Y^{\star} = k )  \,d \bmu_h \,d \bS_h \, d\bx_\star^{m_{\star}} \\ 
    & \approx \frac{1}{M} \sum_{j = 1}^M  \sum_{h_j} \pi_{h_j}^k \int   P( \bx_\star^{o_{\star}}, \bx_\star^{m_{\star}} \, | \bmu_{h_j}, \bS_{h_j},   Y^{\star} = k)  \, d\bx_\star^{m_{\star}} \,\,(\because \text{MC integration})\\
    & =  \frac{1}{M} \sum_{j = 1}^M  \sum_{h_j} \pi_{h_j}^k  P( \bx_\star^{o_{\star}} \, | \bmu_{h_j}^{o_{\star}}, \bS_{h_j}^{o_{\star},o_{\star}},   Y^{\star} = k)   \\
    & = \frac{1}{M} \sum_{j = 1}^M  \sum_{h_j} \pi_{h_j}^k \cdot \mathcal{N}_{|o_{\star}|} \left(\bx_\star^{o_{\star}} \, | \, [\bmu_{h_j}^{k}]^{o_{\star}}, [\bS_{h_j}^k]^{o_{\star},o_{\star}} \right),
\end{aligned}
\label{eq:predn}
\end{equation}
where $M$ is the number of posterior samples generated by MCMC.

\section{Experiments}\label{sec:experiments}

In this section, we evaluate both the imputation and prediction performance of the DPNB model and competitors using multiple datasets. First, we assess the imputation accuracy of our proposed methods with state-of-the-art imputation techniques in various settings. Second,  we quantitatively compare the prediction accuracy of the DPNB model and benchmark classification algorithms based on multiple incomplete datasets in different conditions. In all experiments we apply various missing rates for the covariates (from 10\% to 60\%) and two missingness scenarios: missing completely at random (MCAR), missing at random (MAR). 

We conduct experiments on four real-life datasets from UCI Machine Learning Repository \cite{Dua:2019}: Ecoli, Wine, Breast Cancer Wisconsin (Diagnostic), and Wine Quality datasets. Specifically, the Ecoli dataset contains 336 observations with 8 features and multiple classes. We transform multi-class labels into binary labels which have positive (type im) and negative (the rest). Two discrete variables, Lip and Chg are also removed. The red wine quality dataset contains 1599 observations on 11 attributes of wine and 6 wine quality classes. We divide them into 3 groups for quality: Excellent $(\geq 7)$, Good $(6)$, and Poor $(\leq 5)$. The rest of the multiple datasets are originally used in this experiment. Those datasets have only continuous input variables. The summary of UCI datasets is given in \autoref{tab:data}. Here, the imbalanced ratio (IR) is defined by
\begin{equation*}
    \text{IR} = \dfrac{\max_{C \in \mathcal{A}} |C| }{\min_{C \in \mathcal{A}} |C|},
\end{equation*}
where $\mathcal{A}$ is the set of all classes.  The higher the imbalance ratio is, the bigger disproportion
exists between majority class and minority class.

\begin{table}[ht!]
\centering
	\scriptsize{
\begin{tabular}{cccccc}
	\toprule
 Dataset & \# Samples & \# Features & \# Classes & Class distribution & IR  \\
	\hline
	Ecoli & 336 & 5 &  2 & (259/77) & 3.4\\
	Breast Cancer Wisconsin (Diagnostic) & 569 & 30 & 2 & (357/212)& 1.7\\
	%Parkinsons & 195 & 22 & 2 & (147/48)&  3.1\\
	Wine & 178 & 13 & 3 & (59/71/48)& 1.5\\
	%Seeds & 210 & 7 & 3 & (70/70/70)& 1\\
	Wine Quality & 1599 & 11 & 3 & (744/638/217)&3.4\\
	 \bottomrule
\end{tabular}
}
\caption{Detailed information of UCI datasets.}
\label{tab:data}
\end{table}

\subsection{Imputation performance}\label{subsec:imp}
For each UCI dataset, we generate 100 different incomplete datasets removing 10\% to 60\% of the complete values based on the MCAR or MAR missing assumptions. We make use of the normalized root mean squared errors (NRMSE) as the imputation accuracy measure along with its standard deviation across the 100 replicated datasets. The NRMSE is defined as 
$$
\text{NRMSE} = \sqrt{\dfrac{\text{mean}((X^{\text{true}} - X^{\text{imp}})^2)}{\text{Var}(X^{\text{true}})}},
$$
where $X^{\text{true}}$ is the original data and $X^{\text{imp}}$ the imputed data.

We compare our proposed methods with several popular imputation methods such as multivariate imputation by chained equations using the predictive mean matching; \cite{buuren2010mice} (denoted by MICE), random forest based imputation; \cite{stekhoven2012missforest} (denoted by RF), KNN based imputation with the optimal number of neighbors which minimizes cross-validation errors; \cite{troyanskaya2001missing} (denoted by KNN), imputer utilizing deep denoising autoencoders; \cite{gondara2018mida} (denoted by MIDA), and imputation technique by adapting generative adversarial nets; \cite{yoon2018gain} (denoted by GAIN). 
% We used both R and Python to implement RF (R package missForest \citep{stekhoven2014rpack}), MICE (R package MICE \citep{buuren2010mice}), KNN (R package VIM \citep{kowarik2016imputation}), MIDA () , and GAIN ().
%We use tensorflow to implement GAIN and auto-encoder. 

\begin{figure}[ht!]
  \centering
  \includegraphics[width=\textwidth]{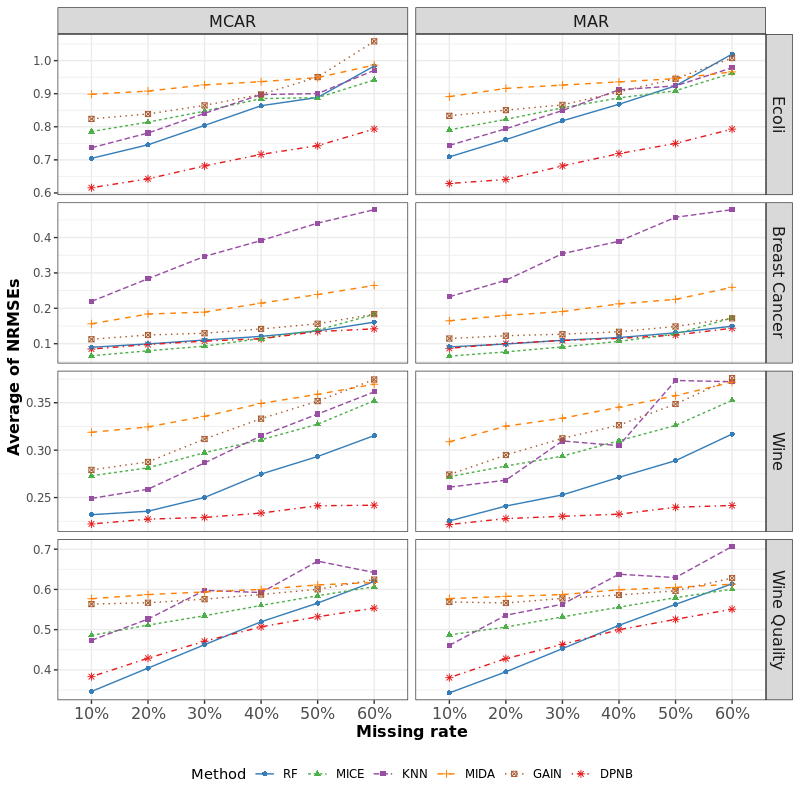}
    \caption{Imputation error for DPNB and competitive models on four UCI datasets with different missing assumptions and missing rates. Curves represent the average of NRMSEs for 100 randomly generated missingness datasets from the four UCI datasets.}
    \label{fig:imp_res}
 \end{figure}

% 결과 분석 (표, 그림)
\autoref{fig:imp_res} shows that the imputer of the DPNB model performs well in most datasets with either the lowest or the second lowest average NRMSE values across 100 replicates irrespective of the missingness mechanism (see \autoref{tab:imp_result} of \autoref{sec:appendB} for details). In particular, we find out from both \autoref{tab:wbcd_mcar_imp} and \autoref{tab:red_mar_imp} that the proposed model is more accurate than other imputation algorithms as missing rates increase. The DPNB model can fill missing values via  different covariance structures constructed by all available observations. However, methods to form conditional distributions using RF and MICE are less accurate than the DPNB because they utilize partial observations and variables instead of full information.  Deep learning-based imputation methods such as GAIN and MIDA have poor performance due to model complexity relative to the number of observed data points. Among all models, the worst-performing method is the KNN based imputation. 

\begin{table}[ht!]
 \centering
\adjustbox{max width=\textwidth}{%
\begin{tabular}{c|cccccc}
  \toprule
  \multirow{2}{*}{Method}&  \multicolumn{6}{c}{Missing Rate} \\
  \cmidrule{2-7}
& 10\% & 20\% & 30\% & 40\% & 50\% & 60\% \\ 
  \midrule
RF & 0.09 (0.038) & 0.099 (0.0258) & 0.11 (0.019) & 0.121 (0.0168) & 0.136 (0.0156) & 0.161 (0.0167)\\ 
 MICE & \textbf{0.066 (0.029)} & \textbf{0.08 (0.0217)} & \textbf{0.093 (0.0163)} & 0.115 (0.0181) & 0.138 (0.0194) & 0.183 (0.0173) \\ 
  KNN & 0.219 (0.0356) & 0.284 (0.027) & 0.347 (0.0215) & 0.392 (0.0166) & 0.441 (0.0188) & 0.479 (0.0175) \\ 
   MIDA & 0.156 (0.0346) & 0.184 (0.0304) & 0.189 (0.0245) & 0.215 (0.0223) & 0.239 (0.0271) & 0.265 (0.0221) \\ 
 GAIN & 0.113 (0.0228) & 0.125 (0.0183) & 0.13 (0.0137) & 0.142 (0.0175) & 0.156 (0.0165) & 0.184 (0.0204) \\ 
  DPNB & 0.085 (0.0522) & 0.098 (0.0431) & 0.107 (0.0362) & \textbf{0.114 (0.029)} & \textbf{0.134 (0.0315)} & \textbf{0.142 (0.0211)} \\
    \bottomrule
\end{tabular}}
\caption{Average normalized mean squared errors with estimated standard errors in parentheses from 100 replications for Breast Cancer Wisconsin (Diagnostic) dataset with varying percentage of missing under MCAR. The best method for each data set is given in bold.}
  \label{tab:wbcd_mcar_imp}
\end{table}

\begin{table}[ht!]
 \centering
\adjustbox{max width=\textwidth}{%
\begin{tabular}{c|cccccc}
  \toprule
  \multirow{2}{*}{Method}&  \multicolumn{6}{c}{Missing Rate} \\
  \cmidrule{2-7}
& 10\% & 20\% & 30\% & 40\% & 50\% & 60\% \\ 
  \midrule
RF & \textbf{0.343 (0.0272)} & \textbf{0.395 (0.0244)} & \textbf{0.453 (0.0215)} & 0.511 (0.0195) & 0.563 (0.0216) & 0.614 (0.0226) \\ 
 MICE & 0.487 (0.0285) & 0.506 (0.0192) & 0.532 (0.0156) & 0.556 (0.0144) & 0.579 (0.0128) & 0.601 (0.0123) \\ 
 KNN & 0.461 (0.0413) & 0.536 (0.0285) & 0.563 (0.0192) & 0.638 (0.0188) & 0.629 (0.0148) & 0.707 (0.0221) \\    
MIDA & 0.577 (0.0266) & 0.583 (0.0225) & 0.587 (0.0179) & 0.599 (0.013) & 0.605 (0.011) & 0.613 (0.0106) \\ 
 GAIN & 0.569 (0.0537) & 0.566 (0.0443) & 0.578 (0.044) & 0.586 (0.0573) & 0.597 (0.0562) & 0.628 (0.0462) \\ 
 DPNB & 0.381 (0.0277) & 0.428 (0.0271) & 0.463 (0.0207) & \textbf{0.5 (0.0186)} & \textbf{0.526 (0.0182)} & \textbf{0.551 (0.0146)} \\
 \bottomrule
\end{tabular}}
\caption{Average normalized mean squared errors with estimated standard errors in parentheses from 100 replications for Wine Quality dataset with varying percentage of missing under MAR. The best method for each data set is given in bold.}
  \label{tab:red_mar_imp}
\end{table}

\subsection{Predictive performance}
We perform 10-fold cross-validation to obtain an estimate of the classification accuracy using simulated missing datasets generated from the previous \autoref{subsec:imp}. The cross-validation process is repeated 10 times for fair comparisons. Since four UCI datasets have various class distributions and binary or multi-class classification problems, we use three types of classification performance metrics to measure the performance of classifiers:(a) accuracy rate, (b) area under the ROC curve (AUC), and (c) F1 score. The classification metrics corresponding to datasets are also shown in \autoref{tab:metric}.

\begin{table}[ht!]
\centering
\footnotesize{
\begin{tabular}{cccccc}
	\toprule
  Ecoli & Breast Cancer & Wine  & Wine Quality \\
	\hline
	F1 score & AUC  & Accuracy rate & F1 score\\
	 \bottomrule
\end{tabular}
}
\caption{Classification performance measures of four UCI datasets.}
\label{tab:metric}
\end{table}

We compare the DPNB model with other prediction models on the incomplete dataset. Other approaches predict test cases after the imputation stage. Some details of the procedures regarding post-imputation prediction are as follows. First, we divide the dataset with missing values into training and test sets. Imputation algorithms utilized in \autoref{subsec:imp} fill missing values of training sets. Second, we build the support vector machines (SVM) with radial basis function (RBF) kernels as a benchmark classifier on the imputed training datasets. Third, missing values in test sets are replaced with the mean of the features of the imputed training set. Finally, we make predictions on the imputed test set using the trained model. We call all competing methods types of imputation algorithms used in both training and test phases.

%%% 결과 분석 표 그림

\begin{figure}[ht!]
  \centering
  \includegraphics[width=\textwidth]{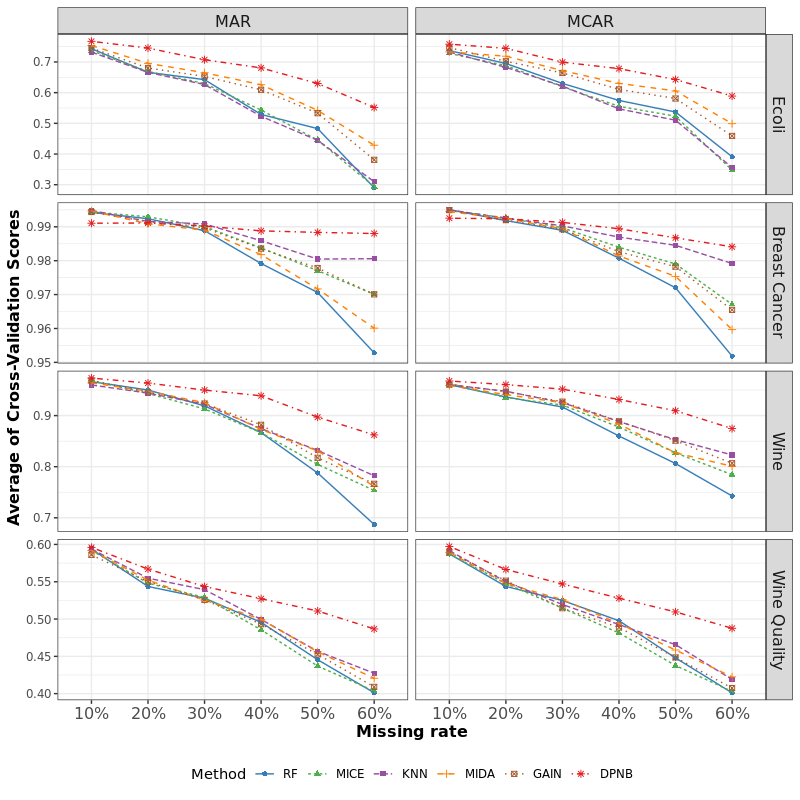}
    \caption{Classification performance for DPNB and competitive models on four UCI datasets with different missing assumptions and missing rates. Curves represent the average of 10 times repeated cross-validation scores such as accuracy, AUC, and F1-score obtained from each method.}
    \label{fig:cl_res}
 \end{figure}

\autoref{fig:cl_res} provides that the DPNB model comes up with the best performance except for the Breast Cancer Wisconsin (Diagnostic) dataset, where it is competitive.  Both \autoref{tab:wbcd_mcar_cl} and \autoref{tab:wbcd_mar_cl} also show that the DPNB has a better classification performance than the others on a dataset for higher missing rates. As missing rates increase, the prediction accuracy of the DPNB is not greatly reduced unlike other competitive approaches. This supports that the DPNB is less affected by both missing data mechanisms and proportions missing proportions than other methods. See \autoref{tab:cl_result} of  \autoref{sec:appendB} for further details about these results.

\begin{table}[ht!]
 \centering
\adjustbox{max width=\textwidth}{%
\begin{tabular}{c|cccccc}
  \toprule
  \multirow{2}{*}{Method}&  \multicolumn{6}{c}{Missing Rate} \\
  \cmidrule{2-7}
& 10\% & 20\% & 30\% & 40\% & 50\% & 60\% \\ 
  \midrule
RF & 0.995 (0.0011) & 0.992 (0.0016) & 0.989 (0.0032) & 0.981 (0.0057) & 0.972 (0.007) & 0.952 (0.0077) \\ 
  MICE & 0.995 (0.0011) & 0.993 (0.002) & 0.99 (0.0024) & 0.984 (0.0059) & 0.979 (0.0079) & 0.967 (0.0041) \\ 
  KNN & \textbf{0.995 (8e-04)} & \textbf{0.993 (0.0017)} & 0.99 (0.0019) & 0.987 (0.0044) & 0.985 (0.0029) & 0.979 (0.0038) \\ 
  MIDA & 0.994 (9e-04) & 0.992 (0.0025) & 0.989 (0.0029) & 0.981 (0.0056) & 0.975 (0.0074) & 0.96 (0.0057) \\ 
  GAIN & 0.995 (0.0013) & 0.992 (0.0023) & 0.989 (0.002) & 0.983 (0.0059) & 0.978 (0.0052) & 0.965 (0.0059) \\ 
  DPNB & 0.993 (0.0025) & 0.992 (0.0029) & \textbf{0.991 (0.002)} & \textbf{0.989 (0.0027)} & \textbf{0.987 (0.0037)} & \textbf{0.984 (0.004)} \\
 \bottomrule
\end{tabular}}
\caption{Mean and standard deviation (in parentheses) for the average AUCs of 10-fold cross-validations on Breast Cancer Wisconsin (Diagnostic) dataset with varying missing rate under MCAR. The best method for each experiment is given in bold.}
  \label{tab:wbcd_mcar_cl}
\end{table}

\begin{table}[ht!]
 \centering
\adjustbox{max width=\textwidth}{%
\begin{tabular}{c|cccccc}
  \toprule
  \multirow{2}{*}{Method}&  \multicolumn{6}{c}{Missing Rate} \\
  \cmidrule{2-7}
& 10\% & 20\% & 30\% & 40\% & 50\% & 60\% \\ 
  \midrule
RF & 0.994 (0.0017) & 0.992 (0.0015) & 0.989 (0.0028) & 0.979 (0.005) & 0.971 (0.0069) & 0.953 (0.0083) \\ 
  MICE & 0.994 (0.0012) & \textbf{0.993 (0.001)} & 0.99 (0.0028) & 0.984 (0.0042) & 0.977 (0.007) & 0.97 (0.0055)  \\ 
  KNN & \textbf{0.995 (0.0015)} & 0.992 (0.0013) & \textbf{0.991 (0.0022)} & 0.986 (0.0047) & 0.98 (0.0052) & 0.981 (0.0057) \\ 
  MIDA & 0.995 (0.0015) & 0.991 (0.0019) & 0.989 (0.0025) & 0.982 (0.0037) & 0.972 (0.0061) & 0.96 (0.0127) \\ 
  GAIN & 0.994 (0.002) & 0.991 (0.0014) & 0.99 (0.0031) & 0.984 (0.0039) & 0.978 (0.0062) & 0.97 (0.0068) \\ 
  DPNB & 0.991 (0.0024) & 0.991 (0.0019) & 0.99 (0.0029) & \textbf{0.989 (0.0031)} & \textbf{0.988 (0.0032)} & \textbf{0.988 (0.003)} \\
 \bottomrule
\end{tabular}}
\caption{Mean and standard deviation (in parentheses) for the average AUCs of 10-fold cross-validations on Breast Cancer Wisconsin (Diagnostic) dataset with varying missing rate under MAR. The best method for each experiment is given in bold.}
  \label{tab:wbcd_mar_cl}
\end{table}

%%%%%%%%%%%%%%%%%%%%%%%%%%%%%%%%%%%%%%%%%%%%%%%%%%%%%%%%%%%%%%%%%%%%%%%%%%%%%%%%%%%%%%%%%%%%%%%%%%%%%%%%%

\section{Applications}\label{sec:analysis}

The main goal of our study is to impute and predict the class of given data with high missing rate. We now apply the DP-Naive Bayes model to a semiconductor manufacturing dataset provided by Samsung Electronics' DS division, which consists of manufacturing operation variables and the semiconductor quality variable related to defect rates. This dataset also includes many missing values and aims to reduce defective wafers made up of semiconductor materials. These challenges are expected to further illustrate the features of the DP-Naive Bayes model.

\begin{figure}[ht!]
     \centering
     \begin{subfigure}{0.49\textwidth}
         \centering
         \includegraphics[width=\textwidth]{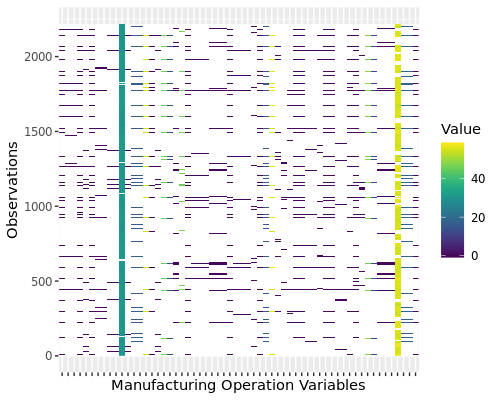}
         \caption{}
         \label{fig:sssc_miss_orig}
     \end{subfigure}
     \begin{subfigure}{0.49\textwidth}
         \centering
         \includegraphics[width=\textwidth]{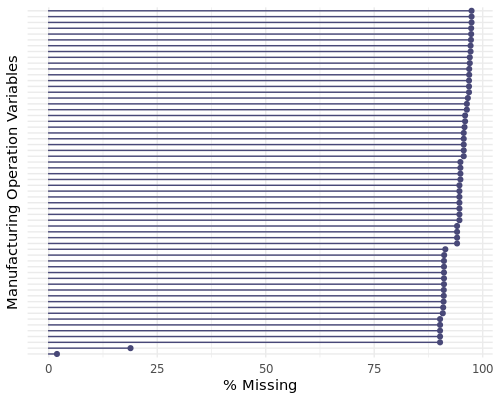}
         \caption{}
         \label{fig:sssc_miss_pct}
     \end{subfigure}
     
        \caption{(a) Missing values (white) of a semiconductor manufacturing dataset, (b) percentage of missing values in each variable.}
        \label{fig:sssc_miss}
\end{figure}

For our analysis, we used the data of 2218 wafer records and 60 manufacturing process variables, discarding features with missing rates larger than 97.5\% and categorical values. \autoref{fig:sssc_miss_orig} depicts the observed and missing values of the pre-processing data in colored tiles and white tiles, respectively. It is made up of mostly white tiles. Missing values in the dataset account for about 95\% of all values and most manufacturing operation variables have over 90\% missing rate, as can be seen in the \autoref{fig:sssc_miss_pct}. We can say that the data is MCAR since missing values seem to be pretty randomly scattered and occur at random in the manufacturing processes. Furthermore, the target metrology variable related to defect rates is highly imbalanced, with most records falling in the “No defect” class. The imbalance ratio of this dataset is  roughly 9.4.

\subsection{Imputation performance} 

To assess the imputation accuracy based on the real world dataset, we consider three scenarios by deleting artificially 50, 100, and 500 complete values in the semiconductor dataset under MCAR assumption. In every cases, we made 100 replicates, respectively.  For performance comparisons, we then compute the average of normalized root mean squared errors obtained by each method in all cases.

\begin{figure}[ht!]
  \centering
  \includegraphics[width=\textwidth]{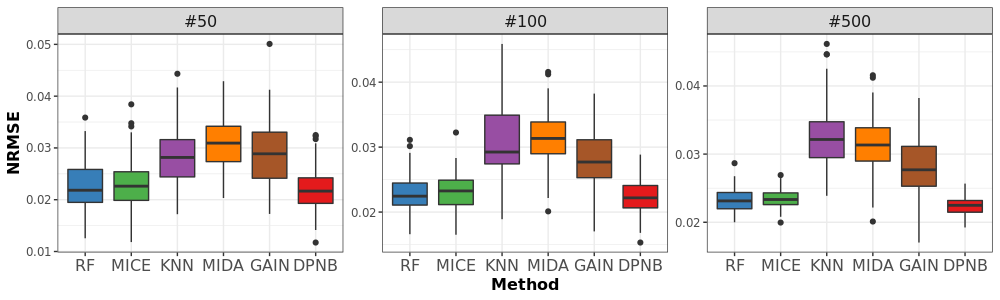}
    \caption{Boxplots of NRMSEs over 100 replications for semiconductor dataset with varying number of artificial missing values.}
    \label{fig:sssc_imp_bp}
 \end{figure}

The results of the DPNB and competing models are provided in \autoref{fig:sssc_imp_bp} and \autoref{tab:sssc_imp_nrmse}. They demonstrate that the DPNB model provides more accurate values and consistent performance through both minimum mean and the lowest standard deviation values over 100 replications of NRMSEs. As  shown in the Breast Cancer example, both RF and MICE yield similar imputation performance with the DPNB model. Imputation methods based on deep architectures including MIDA and GAIN have difficulty selecting appropriate parameters to prevent the problem of overfitting. These methods have then lower imputation accuracy than the DPNB, RF and MICE.

\begin{table}[ht]
\centering
\begin{tabular}{cccc}
  \toprule
  & \#50 & \#100 & \#500 \\ 
  \midrule
  RF & 0.0226 (0.00452) & 0.0226 (0.00296) & 0.0233 (0.00163) \\ 
  MICE & 0.023 (0.00457) & 0.0233 (0.00287) & 0.0234 (0.0014) \\ 
  KNN & 0.0284 (0.00562) & 0.0309 (0.00566) & 0.0325 (0.00442) \\ 
  MIDA & 0.031 (0.00513) & 0.0312 (0.00394) & 0.0312 (0.00394) \\ 
  GAIN & 0.029 (0.00617) & 0.0281 (0.00418) & 0.0281 (0.00418) \\ 
  DPNB & \textbf{0.0219 (0.00423)} & \textbf{0.0222 (0.00282)} & \textbf{0.0224 (0.0013)} \\ 
   \bottomrule
\end{tabular}
\caption{Average of NRMSEs over 100 replications for semiconductor dataset with different numbers of artificial missing values under MCAR. Estimated standard errors of NRMSEs are shown in parentheses. The best method for each experiment is given in bold.
}
\label{tab:sssc_imp_nrmse}
\end{table}

\subsection{Predictive performance} 

 In this experiment, we randomly partitioned the real dataset into training and test sets with different ratios. The training set was set from 50\% to 90\% of the overall dataset and the test set was accordingly set from 50\% to 10\% of it. For example, if 90\% of the data is used as the training set, then the remaining 10\% is used as the test set. Likewise, we created 100 replicate datasets in all cases. Furthermore, the bootstrapping-based oversampling technique that replicates observations from minority class was applied into training sets in order to balance the two classes. Note that the same indices of observations selected from the minority class were used in all methods given percentages of the training sets. 

We add the SVM with a linear kernel as another base classifier. We will name competitive methods by combining imputation techniques and kernels of support vector machine, e.g., ``RF+L" means that the RF imputes missing values of training sets and the SVM with a linear kernel is fitted on the imputed training sets and predicts the test set values. The results of this experiment are given in \autoref{tab:sssc_cls_result}. The columns of competing methods represent their average F1-score over 100 replicate datasets, normalized by the average F1-score of the DPNB model. The values higher than one indicate that the methods provide better performance than the DPNB and values lower than one indicate worse performance in imbalanced settings.

\begin{table}[ht]
\adjustbox{max width=\textwidth}{
\centering
\begin{tabular}{lcccccccccccc}
  \toprule
 & RF+L & RF+R & MICE+L& MICE+R & KNN+L & KNN+R & MIDA+L & MIDA+R & GAIN+L & GAIN+R & DPNB \\
  \midrule
50\% / 50\%  & 0.9516 (3) & 0.6638 (9) & 0.8618 (6) & 0.0349 (11) & 0.8359 (7) & 0.7016 (8) & 0.6508 (10) & 0.9813 (2) & 0.8891 (5) & 0.9461 (4) & \textbf{1 (1)} \\ 
  60\% / 40\% & 0.9829 (3) & 0.7192 (8) & 0.8833 (6) & 0.0282 (11) & 0.8559 (7) & 0.7067 (9) & 0.6434 (10) & 0.9786 (4) & 0.9262 (5) & \textbf{1.0096 (1)} & 1 (2) \\ 
  70\% / 30\% & 0.9845 (5) & 0.7248 (9) & 0.9186 (6) & 0.0427 (11) & 0.8752 (7) & 0.7423 (8) & 0.7174 (10) & \textbf{1.0419 (1)} & 1.0051 (2) & 0.9851 (4) & 1 (3) \\ 
  80\% / 20\% & \textbf{1.0184 (1)} & 0.6591 (10) & 0.9391 (6) & 0.0453 (11) & 0.9019 (7) & 0.7314 (9) & 0.7814 (8) & 0.9965 (4) & 1.0124 (2) & 0.971 (5) & 1 (3) \\ 
  90\% / 10\%  & 0.9982 (2) & 0.6281 (10) & 0.9621 (6) & 0.0298 (11) & 0.8186 (7) & 0.7394 (8) & 0.7329 (9) & 0.9841 (3) & 0.9662 (5) & 0.9825 (4) & \textbf{1 (1)} \\   

& & & & & & & & & & & \\
Average nomalized F1-score & 0.987 & 0.679 & 0.913 & 0.036 & 0.857 & 0.724 & 0.705 & 0.996 & 0.960 & 0.979 & \textbf{1.000}\\ 
Average rank &2.8 & 9.2 & 6.0 & 11.0 & 7.0 & 8.4 & 9.4 & 2.8 & 3.8 & 3.6 & \textbf{2.0} \\
   \bottomrule
\end{tabular}
}
\caption{Normalized average F1-scores over 100 replications for the semiconductor dataset with different ratios between training and test sets. The rank of the methods judged by normalized F1-scores is shown in parentheses. The best method for each experiment is given in bold.}
\label{tab:sssc_cls_result}
\end{table}
As shown in \autoref{tab:sssc_cls_result}, the DPNB method performs well in practice since it ranks high in almost all cases. In particular, it yields the best prediction performance on two out of the five ratios. We can say that the DPNB model provides stable performance by obtaining the highest average normalized F1-score and average rank and stable inference in the incomplete data with high missing rates. However, since the semiconductor dataset has very high missing rates, the DPNB does not provide remarkable performance against other methods as shown in the previous section. The methods using the KNN imputer have poor performance irrespective of the type of classifiers. We also present the actual average F1-score in \autoref{tab:sssc_cls_f1score} of  \autoref{sec:appendB} for more details.

%%%%%%%%%%%%%%%%%%%%%%%%%%%%%%%%%%%%%%%%%%%%%%%%%%%%%%%%%%%%%%%%%%%%%%%%%%%%%%%%%%%%%%%%%%%%%%%%%%%%%%%%%

\section{Conclusions}\label{sec:discussion}

We propose a new method for classification problems with incomplete data based on the Dirichlet process and naive Bayes model, DPNB. The DPNB method is free from the distribution assumption and constructs flexible imputer and classifier. The flexibility and effect of the DPNB model are verified by various experiments. Moreover, the DPNB model suffers less from the overfitting problem, which frequently occurs using a flexible model, by considering proper prior. The DPNB model shows stable and better performance than other methods on experiments even missing rate is high. As an improvement to the DPNB model, we would like to address the problem with computational time. Due to the limit of the MCMC algorithm, the DPNB model takes a longer time compared to other models. In future studies, we would like to propose ways to reduce computational time for the DPNB model such as variational methods.

\section*{Acknowledgement}
This work was supported by Samsung Electronics Co., Ltd. (IO210216-08417-01). 
\bibliographystyle{dcu}
\bibliography{DPNB.bib}

\newpage

\appendix
\begin{appendices}
\section{Slice sampling for the DPMM on incomplete data} \label{sec:appendA}
Based on the model \eqref{eq:dpmm}, approximate posterior of $\bX$ conditioned on the class $k$ is computed using an improved slice sampler for the DPMM presented by \cite{ge2015distributed}. However, since observations $\textbf{x}_i :=  \{\textbf{x}_i^{o_i}, \textbf{x}_i^{m_i}\},\,  i = 1, \ldots, n$ contain missing values, we additionally embed updating steps \eqref{eq:normalcond} for estimating the missing values in the slice sampling algorithm for the DPMM. We can summarize the MCMC algorithm for $\bX$ conditioned on the class $k$ as follows:

\bigskip

\noindent \textbf{[Step 1]} For each cluster $h$, sample mixing proportions $\textbf{w}^k := (w_1^k, w^k_2, \ldots, w^k_H, w_*^k)$ 
$$
[\,\textbf{w}^k \,|\, \text{others} \,] \sim \text{Dir}(\pi_1^k, \ldots, \pi_H^k, \alpha),
$$
where $\pi_h^k := |\{i: z_i^k = h\}|$ and $H$ is the number of (current) clusters.

\bigskip 

\noindent \textbf{[Step 2]} Sample auxilrary variables and set the minimum
\begin{gather*}
[ \, u_i^k \,|\, \text{others} \,]\sim \mathcal{U}(0, w^k_{z^k_i}), \quad \forall i = 1, \ldots, n_k, \\
u_{*}^k = \min_i u_i^k.
\end{gather*}

\bigskip 

\noindent \textbf{[Step 3]} Generate new clusters through stick-breaking processes until $w_*^k < u_*^k$
\begin{gather*}
H^k \leftarrow H^k + 1, \quad  \nu^k_{H^k} \sim \text{Beta}(1, \alpha), \\
w^k_{H^k} = w^k_* \times  \nu^k_{H^k},\quad w_*^k \leftarrow w_*^k \times (1-\nu^k_{H^k}), \\
(\bmu^k_{H^k},\bS^k_{H^k})  \sim \mathcal{N}(\bmu^k; \mathbf{m}_0^k, \bS^k/\tau_0^k) \cdot \mathcal{I}\mathcal{W}(\bS^k; \mathbf{B}_0^k,\nu_0^k),
\end{gather*}
where $w_*^k$ is the remaining stick length.

\bigskip 

\noindent \textbf{[Step 4]} For each observation $\mathbf{x}^k_i$, sample the assignment variable $z^k_i$
\begin{gather*}
    p(z_i = h \,|\, \text{others}) \propto I(w_h^k \geq u^k_i)  \cdot \mathcal{N}(\bx_i^k; \bmu_i^k, \bS_i^k).
\end{gather*}
for $h = 1, \ldots, H^k$.

\bigskip 

\noindent \textbf{[Step 5]} For each cluster $h$, sample cluster parameters $(\bmu^k_{h},\bS^k_{h})$
\begin{gather*}
    p(\bmu^k_{h},\bS^k_{h} \,|\, \text{others}) \propto \mathcal{N}(\bmu^k_h; \mathbf{m}_0^k, \bS^k/\tau_0^k) \cdot \mathcal{I}\mathcal{W}(\bS^k_h; \mathbf{B}_0^k,\nu_0^k) \prod_{\{i:\, z_i^k = h\}} \mathcal{N}(\bx_i^k; \bmu_i^k, \bS_i^k).
\end{gather*}

\bigskip

\noindent \textbf{[Step 6]} For each observation $\mathbf{x}_i^k$ that belongs  to specific $h$th  mixure  component, sample the missing values $\mathbf{x}_i^{m_i}$
\begin{gather*}
[\,\bx_i^{m_i} \,|\, \text{others}\,] \sim \mathcal{N}_{|m_i|}((\bmu_h^k)^{m_i|o_i}, (\bS_h^k)^{m_i | o_i}), \\
(\bmu_h^k)^{m_i|o_i} = (\bmu_h^k)^{m_i} + (\bS_h^k)^{m_i, o_i} ((\bS_h^k)^{o_i, o_i})^{-1}( \bx_i^{o_i} - (\bmu_h^k)^{o_i}), \\
(\bS_h^k)^{m_i|o_i} = (\bS_h^k)^{m_i, m_i} - (\bS_h^k)^{m_i, o_i}((\bS_h^k)^{o_i, o_i})^{-1}(\bS_h^k)^{o_i, m_i}. 
\end{gather*}

\bigskip

\noindent \textbf{[Step 7]} Repeat steps 1-6 until convergence.

\newpage

\section{Full simulation results} \label{sec:appendB}

In this section we present additional simulation results for imputation and predictive performance reported in \autoref{sec:experiments} and \autoref{sec:analysis}. 
% \autoref{tab:imp_result} presents average of NRMSEs for the state-of-the-art imputation methods and the DPNB model in four UCI datasets with different missingness patterns (MCAR and MAR) and missing rates (10\% to  60\%). Similarly, \autoref{tab:cl_result} provides  average of 10 times repeated 10-folds cross-validated metrics including accuracy, AUC, and F1-score. \autoref{tab:sssc_cls_f1score} shows average of F1-scores over 100 replicated training and test sets generated from the semiconductor dataset with different split ratios (50\% to 90\%). 

%%%%%%%%%%%%%%%%%%%%%%%%%%%%%%%%%%%%%%%%%%%%%%%%%%%%%%%%%%%%%
%%%%%%%%%%%%%%%%%%%%%%%%%%%%%%%%%%%%%%%%%%%%%%%%%%%%%%%%%%%%%

\begin{table}[ht!]
 \centering
\adjustbox{max width=\textheight, angle = 90, max height = 1.09\textwidth }{%
 \footnotesize
	\begin{tabular}{c|c|cccccc|cccccc}
		\toprule
		\multirow{2}{*}{Dataset}& 	\multirow{2}{*}{Model}&  \multicolumn{6}{c|}{MCAR}& \multicolumn{6}{c}{MAR}\\
		\cmidrule{3-14}

	&	& 10\% & 20\% & 30\% & 40\% & 50\%& 60\% & 10\% & 20\% & 30\% & 40\% & 50\%& 60\% \\
		\midrule
\multirow{6}{*}{Ecoli} & RF & 0.705 (0.0564) & 0.746 (0.0398) & 0.804 (0.0341) & 0.864 (0.0391) & 0.889 (0.0323) & 0.984 (0.0367) & 0.709 (0.055) & 0.761 (0.0467) & 0.818 (0.0385) & 0.868 (0.0345) & 0.924 (0.0463) & 1.02 (0.0857) \\ 
   & MICE & 0.786 (0.0589) & 0.814 (0.0346) & 0.847 (0.0279) & 0.885 (0.0229) & 0.888 (0.022) & 0.941 (0.0225) & 0.791 (0.055) & 0.822 (0.0371) & 0.857 (0.0297) & 0.887 (0.0244) & 0.909 (0.0385) & 0.962 (0.0598)\\ 
  & KNN & 0.737 (0.0619) & 0.781 (0.0427) & 0.841 (0.0386) & 0.897 (0.0369) & 0.9 (0.0339) & 0.971 (0.0393) & 0.744 (0.0553) & 0.794 (0.0383) & 0.849 (0.0343) & 0.912 (0.0319) & 0.924 (0.0486) & 0.978 (0.0851) \\ 
& MIDA & 0.898 (0.0416) & 0.908 (0.0415) & 0.927 (0.0295) & 0.936 (0.0229) & 0.949 (0.0243) & 0.987 (0.0247) & 0.891 (0.0429) & 0.916 (0.0431) & 0.926 (0.0287) & 0.936 (0.0233) & 0.946 (0.0269) & 0.966 (0.0259) \\ 
& GAIN & 0.824 (0.0675) & 0.839 (0.052) & 0.865 (0.0477) & 0.898 (0.0489) & 0.95 (0.0801) & 1.058 (0.1221) & 0.833 (0.068) & 0.85 (0.0498) & 0.866 (0.0402) & 0.906 (0.0565) & 0.945 (0.0784) & 1.008 (0.0888) \\ 

& DPNB & 0.616 (0.0521) & 0.643 (0.0422) & 0.682 (0.0383) & 0.717 (0.0265) & 0.743 (0.0287) & 0.793 (0.0256) & 0.629 (0.0481) & 0.641 (0.0425) & 0.682 (0.0327) & 0.719 (0.0268) & 0.75 (0.0278) & 0.793 (0.0253)  \\  

		\midrule
\multirow{6}{*}{Breast Cancer}  & RF & 0.09 (0.038) & 0.099 (0.0258) & 0.11 (0.019) & 0.121 (0.0168) & 0.136 (0.0156) & 0.161 (0.0167) & 0.091 (0.03) & 0.099 (0.024) & 0.11 (0.0183) & 0.117 (0.0174) & 0.131 (0.016) & 0.15 (0.0143) \\ 
  & MICE & 0.066 (0.029) & 0.08 (0.0217) & 0.093 (0.0163) & 0.115 (0.0181) & 0.138 (0.0194) & 0.183 (0.0173) & 0.065 (0.0236) & 0.077 (0.0206) & 0.091 (0.017) & 0.106 (0.0162) & 0.127 (0.0176) &  0.173 (0.0158) \\ 
  & KNN & 0.219 (0.0356) & 0.284 (0.027) & 0.347 (0.0215) & 0.392 (0.0166) & 0.441 (0.0188) & 0.479 (0.0175) & 0.233 (0.0357) & 0.279 (0.0261) & 0.355 (0.0199) & 0.389 (0.0206) & 0.457 (0.0178) & 0.479 (0.0169) \\ 
&  MIDA & 0.156 (0.0346) & 0.184 (0.0304) & 0.189 (0.0245) & 0.215 (0.0223) & 0.239 (0.0271) & 0.265 (0.0221) & 0.165 (0.0351) & 0.18 (0.0305) & 0.191 (0.0245) & 0.213 (0.0243) & 0.225 (0.0223) & 0.259 (0.0225) \\
& GAIN & 0.113 (0.0228) & 0.125 (0.0183) & 0.13 (0.0137) & 0.142 (0.0175) & 0.156 (0.0165) & 0.184 (0.0204) & 0.115 (0.0235) & 0.122 (0.0164) & 0.127 (0.0141) & 0.134 (0.013) & 0.149 (0.0159) & 0.172 (0.0195) \\ 
 
& DPNB & 0.085 (0.0522) & 0.098 (0.0431) & 0.107 (0.0362) & 0.114 (0.029) & 0.134 (0.0315) & 0.142 (0.0211) & 0.087 (0.0585) & 0.101 (0.0464) & 0.109 (0.0384) & 0.115 (0.0298) & 0.124 (0.0291) & 0.144 (0.0208) \\ 

% 		\midrule
% \multirow{5}{*}{Parkinsons}  & RF & 0.306 (0.0914) & 0.328 (0.0647) & 0.327 (0.0438) & 0.33 (0.0373) & 0.35 (0.0289) & 0.367 (0.0223) & 0.298 (0.094) & 0.315 (0.0639) & 0.332 (0.0467) & 0.327 (0.0351) & 0.34 (0.0294) & 0.329 (0.0349) \\ 
%   & MICE & NA & NA & NA & NA & NA & NA & NA &NA  & NA & NA & NA & NA  \\ 
%   & KNN & 0.368 (0.0832) & 0.4 (0.0604) & 0.403 (0.0387) & 0.41 (0.0322) & 0.414 (0.0254) & 0.423 (0.0184) & 0.358 (0.0812) & 0.392 (0.0575) & 0.41 (0.0374) & 0.407 (0.0311) & 0.415 (0.0236) & 0.401 (0.0282) \\ 
% & DPNB1 & 0.737 (0.0619) & 0.781 (0.0427) & 0.841 (0.0386) & 0.737 (0.0619) & 0.781 (0.0427) & 0.841 (0.0386) & 0.744 (0.0553) & 0.794 (0.0383) & 0.849 (0.0343) & 0.744 (0.0553) & 0.794 (0.0383) & 0.849 (0.0343) \\ 
% & DPNB & 0.318 (0.0968) & 0.344 (0.0674) & 0.346 (0.0466) & 0.347 (0.039) & 0.357 (0.0289) & 0.372 (0.0243) & 0.318 (0.092) & 0.329 (0.0658) & 0.346 (0.0441) & 0.344 (0.0359) & 0.354 (0.0322) & 0.341 (0.0323)  \\ 

		\midrule
\multirow{6}{*}{Wine} & RF & 0.232 (0.0361) & 0.236 (0.0326) & 0.25 (0.0293) & 0.275 (0.027) & 0.293 (0.0236) & 0.315 (0.025) & 0.225 (0.036) & 0.241 (0.0315) & 0.253 (0.0288) & 0.271 (0.0259) & 0.289 (0.0209) & 0.317 (0.0229) \\ 
  & MICE & 0.273 (0.0521) & 0.281 (0.0352) & 0.297 (0.031) & 0.311 (0.0275) & 0.327 (0.023) & 0.352 (0.018) & 0.272 (0.0456) & 0.283 (0.0348) & 0.294 (0.0304) & 0.31 (0.0291) & 0.326 (0.0243) & 0.353 (0.0234) \\ 
  & KNN & 0.249 (0.052) & 0.259 (0.0486) & 0.287 (0.04) & 0.315 (0.0358) & 0.338 (0.038) & 0.361 (0.0409) & 0.261 (0.0576) & 0.268 (0.0446) & 0.31 (0.0416) & 0.305 (0.0338) & 0.373 (0.0363) & 0.372 (0.0368) \\ 
  
& MIDA & 0.319 (0.0604) & 0.324 (0.039) & 0.336 (0.0321) & 0.349 (0.0257) & 0.359 (0.0228) & 0.37 (0.0208) & 0.309 (0.0511) & 0.325 (0.0345) & 0.334 (0.0287) & 0.345 (0.0256) & 0.357 (0.0253) & 0.372 (0.024) \\ 

& GAIN & 0.279 (0.0382) & 0.288 (0.0317) & 0.312 (0.0293) & 0.333 (0.0354) & 0.352 (0.032) & 0.375 (0.0312) & 0.274 (0.0429) & 0.295 (0.0332) & 0.312 (0.0293) & 0.326 (0.0315) & 0.349 (0.0286) & 0.376 (0.0298) \\
 
& DPNB & 0.222 (0.0391) & 0.227 (0.0305) & 0.229 (0.0237) & 0.234 (0.0166) & 0.241 (0.0185) & 0.242 (0.0169) & 0.222 (0.0413) & 0.228 (0.0284) & 0.23 (0.0217) & 0.232 (0.0216) & 0.24 (0.0178) & 0.242 (0.0187) \\ 

% 		\midrule
% \multirow{5}{*}{Seeds}  & RF & 0.101 (0.0205) & 0.106 (0.0128) & 0.109 (0.0102) & 0.117 (0.0086) & 0.13 (0.01) & 0.143 (0.0104) & 0.099 (0.0185) & 0.102 (0.0125) & 0.107 (0.009) & 0.109 (0.0079) & 0.116 (0.0086) & 0.131 (0.0088) \\ 
%   & MICE & 0.108 (0.022) & 0.111 (0.0135) & 0.115 (0.0109) & 0.122 (0.0092) & 0.136 (0.0089) & 0.153 (0.0091) & 0.106 (0.0202) & 0.107 (0.0117) & 0.11 (0.009) & 0.113 (0.0083) & NA & NA \\ 
%   & KNN & 0.124 (0.0222) & 0.149 (0.0167) & 0.163 (0.0136) & 0.191 (0.0129) & 0.214 (0.0128) & 0.236 (0.0152) & 0.12 (0.0175) & 0.133 (0.013) & 0.155 (0.0117) & 0.181 (0.0117) & 0.218 (0.0174) & 0.243 (0.0125) \\ 
% & DPNB1 & 0.737 (0.0619) & 0.781 (0.0427) & 0.841 (0.0386) & 0.737 (0.0619) & 0.781 (0.0427) & 0.841 (0.0386) & 0.744 (0.0553) & 0.794 (0.0383) & 0.849 (0.0343) & 0.744 (0.0553) & 0.794 (0.0383) & 0.849 (0.0343) \\ 
% & DPNB & 0.092 (0.0204) & 0.094 (0.0115) & 0.094 (0.0087) & 0.097 (0.008) & 0.101 (0.0065) & 0.107 (0.0065) & 0.091 (0.0156) & 0.092 (0.0106) & 0.094 (0.0079) & 0.096 (0.0066) & 0.099 (0.0064) & 0.106 (0.0064)
%  \\ 

		\midrule
\multirow{6}{*}{Wine Quality} & RF & 0.346 (0.0263) & 0.404 (0.0258) & 0.463 (0.0225) & 0.52 (0.0243) & 0.566 (0.0216) & 0.62 (0.0224) & 0.343 (0.0272) & 0.395 (0.0244) & 0.453 (0.0215) & 0.511 (0.0195) & 0.563 (0.0216) & 0.614 (0.0226) \\ 
  & MICE & 0.486 (0.031) & 0.511 (0.0193) & 0.535 (0.018) & 0.56 (0.0136) & 0.584 (0.0148) & 0.607 (0.0097) & 0.487 (0.0285) & 0.506 (0.0192) & 0.532 (0.0156) & 0.556 (0.0144) & 0.579 (0.0128) & 0.601 (0.0123) \\ 
   & KNN & 0.473 (0.0377) & 0.526 (0.0291) & 0.598 (0.0207) & 0.592 (0.0167) & 0.67 (0.0195) & 0.642 (0.0134) & 0.461 (0.0413) & 0.536 (0.0285) & 0.563 (0.0192) & 0.638 (0.0188) & 0.629 (0.0148) & 0.707 (0.0221) \\ 
& MIDA & 0.577 (0.027) & 0.587 (0.0209) & 0.594 (0.0166) & 0.6 (0.0143) & 0.611 (0.012) & 0.617 (0.0102) & 0.577 (0.0266) & 0.583 (0.0225) & 0.587 (0.0179) & 0.599 (0.013) & 0.605 (0.011) & 0.613 (0.0106) \\ 
& GAIN & 0.564 (0.0562) & 0.567 (0.0457) & 0.576 (0.0438) & 0.587 (0.0423) & 0.6 (0.0443) & 0.624 (0.0477) & 0.569 (0.0537) & 0.566 (0.0443) & 0.578 (0.044) & 0.586 (0.0573) & 0.597 (0.0562) & 0.628 (0.0462) \\
& DPNB & 0.383 (0.029) & 0.429 (0.0311) & 0.471 (0.0267) & 0.507 (0.022) & 0.532 (0.0189) & 0.553 (0.0149) & 0.381 (0.0277) & 0.428 (0.0271) & 0.463 (0.0207) & 0.5 (0.0186) & 0.526 (0.0182) & 0.551 (0.0146) \\ 

 \bottomrule
	\end{tabular}}
  \caption{Average of NRMSEs over 100 replications for four UCI datasets with different missingness patterns and missing rates. Estimated standard errors of NRMSEs are shown in parentheses.}
  \label{tab:imp_result}
\end{table}

\clearpage
%%%%%%%%%%%%%%%%%%%%%%%%%%%%%%%%%%%%%%%%%%%%%%%%%%%%%%%%%%%%%
%%%%%%%%%%%%%%%%%%%%%%%%%%%%%%%%%%%%%%%%%%%%%%%%%%%%%%%%%%%%%

\begin{table}[ht!]
 \centering
\adjustbox{max width=0.9\textheight, max height = \textwidth, angle = 90}{%
 \footnotesize
	\begin{tabular}{c|c|c|cccccc|cccccc}
		\toprule
		\multirow{2}{*}{Dataset} & \multirow{2}{*}{Metric}& 	\multirow{2}{*}{Model}&  \multicolumn{6}{c|}{MCAR}& \multicolumn{6}{c}{MAR}\\
		\cmidrule{4-15}

	& &	& 10\% & 20\% & 30\% & 40\% & 50\%& 60\% & 10\% & 20\% & 30\% & 40\% & 50\%& 60\% \\
		\midrule
\multirow{7}{*}{Ecoli} & \multirow{7}{*}{F1 score} &  RF & 0.737 (0.027) & 0.695 (0.0304) & 0.63 (0.0378) & 0.575 (0.0777) & 0.537 (0.0516) & 0.391 (0.065) & 0.743 (0.0286) & 0.666 (0.0371) & 0.642 (0.0418) & 0.529 (0.0747) & 0.483 (0.0656) & 0.29 (0.0859) \\ 
  & &  MICE & 0.728 (0.0268) & 0.688 (0.0308) & 0.619 (0.0337) & 0.556 (0.0577) & 0.523 (0.0565) & 0.348 (0.0627) & 0.737 (0.029) & 0.669 (0.0296) & 0.63 (0.0548) & 0.543 (0.0605) & 0.446 (0.0758) & 0.294 (0.0826) \\ 
  & &  KNN & 0.732 (0.0334) & 0.682 (0.0288) & 0.621 (0.0262) & 0.547 (0.0587) & 0.51 (0.0399) & 0.356 (0.1007) & 0.732 (0.0284) & 0.666 (0.0434) & 0.626 (0.0322) & 0.523 (0.0628) & 0.445 (0.095) & 0.31 (0.0815) \\ 
  & &  MIDA & 0.733 (0.0309) & 0.719 (0.0498) & 0.672 (0.0291) & 0.63 (0.0475) & 0.606 (0.0408) & 0.499 (0.0483) & 0.754 (0.0237) & 0.696 (0.0301) & 0.665 (0.0387) & 0.626 (0.0377) & 0.542 (0.0463) & 0.429 (0.067) \\ 
  & & GAIN & 0.747 (0.0248) & 0.702 (0.0365) & 0.664 (0.016) & 0.611 (0.0571) & 0.581 (0.0361) & 0.459 (0.072) & 0.745 (0.0237) & 0.68 (0.0376) & 0.653 (0.0379) & 0.609 (0.0609) & 0.533 (0.0725) & 0.381 (0.0545) \\ 
  & & DPNB & 0.758 (0.0182) & 0.744 (0.0302) & 0.699 (0.0225) & 0.678 (0.034) & 0.643 (0.0429) & 0.589 (0.0302) & 0.767 (0.0254) & 0.745 (0.0202) & 0.707 (0.0242) & 0.681 (0.0457) & 0.63 (0.0384) & 0.551 (0.0715) \\ 
%   & & DPNB-T & 0.76 (0.017) & 0.746 (0.0308) & 0.701 (0.0257) & 0.68 (0.0352) & 0.643 (0.0438) & 0.589 (0.0309) & 0.773 (0.0324) & 0.745 (0.0221) & 0.71 (0.0254) & 0.684 (0.0492) & 0.632 (0.0395) & 0.55 (0.0725) \\ 

		\midrule
\multirow{7}{*}{Breast Cancer} & \multirow{7}{*}{AUC} &  RF & 0.995 (0.0011) & 0.992 (0.0016) & 0.989 (0.0032) & 0.981 (0.0057) & 0.972 (0.007) & 0.952 (0.0077) & 0.994 (0.0017) & 0.992 (0.0015) & 0.989 (0.0028) & 0.979 (0.005) & 0.971 (0.0069) & 0.953 (0.0083) \\ 
  & & MICE & 0.995 (0.0011) & 0.993 (0.002) & 0.99 (0.0024) & 0.984 (0.0059) & 0.979 (0.0079) & 0.967 (0.0041) & 0.994 (0.0012) & 0.993 (0.001) & 0.99 (0.0028) & 0.984 (0.0042) & 0.977 (0.007) &  0.97 (0.0055) \\ 
  & & KNN & 0.995 (8e-04) & 0.993 (0.0017) & 0.99 (0.0019) & 0.987 (0.0044) & 0.985 (0.0029) & 0.979 (0.0038) & 0.995 (0.0015) & 0.992 (0.0013) & 0.991 (0.0022) & 0.986 (0.0047) & 0.98 (0.0052) & 0.981 (0.0057) \\ 
  & & MIDA & 0.994 (9e-04) & 0.992 (0.0025) & 0.989 (0.0029) & 0.981 (0.0056) & 0.975 (0.0074) & 0.96 (0.0057) & 0.995 (0.0015) & 0.991 (0.0019) & 0.989 (0.0025) & 0.982 (0.0037) & 0.972 (0.0061) & 0.96 (0.0127) \\ 
  & & GAIN & 0.995 (0.0013) & 0.992 (0.0023) & 0.989 (0.002) & 0.983 (0.0059) & 0.978 (0.0052) & 0.965 (0.0059) & 0.994 (0.002) & 0.991 (0.0014) & 0.99 (0.0031) & 0.984 (0.0039) & 0.978 (0.0062) & 0.97 (0.0068) \\ 
  & & DPNB & 0.993 (0.0025) & 0.992 (0.0029) & 0.991 (0.002) & 0.989 (0.0027) & 0.987 (0.0037) & 0.984 (0.004) & 0.991 (0.0024) & 0.991 (0.0019) & 0.99 (0.0029) & 0.989 (0.0031) & 0.988 (0.0032) & 0.988 (0.003) \\ 
%   & & DPNB-T & 0.993 (0.0013) & 0.993 (0.0011) & 0.993 (0.0015) & 0.99 (0.0025) & 0.988 (0.0033) & 0.985 (0.0041) & 0.992 (0.0017) & 0.992 (0.0016) & 0.991 (0.0032) & 0.989 (0.005) & 0.987 (0.0038) & 0.988 (0.0031) \\  

		\midrule
\multirow{7}{*}{Wine} &  \multirow{7}{*}{Accuracy rate} &  RF & 0.96 (0.0143) & 0.936 (0.0205) & 0.917 (0.0302) & 0.86 (0.0174) & 0.806 (0.0301) & 0.743 (0.019) & 0.966 (0.01) & 0.95 (0.0155) & 0.919 (0.0168) & 0.867 (0.0148) & 0.788 (0.0316) & 0.687 (0.0451) \\ 
  & & MICE & 0.963 (0.0115) & 0.935 (0.0179) & 0.921 (0.0169) & 0.878 (0.0219) & 0.827 (0.0217) & 0.784 (0.0243) & 0.969 (0.0095) & 0.943 (0.0183) & 0.913 (0.0167) & 0.867 (0.0238) & 0.805 (0.0376) &  0.754 (0.02) \\ 
   & & KNN & 0.961 (0.0133) & 0.948 (0.0137) & 0.925 (0.0187) & 0.888 (0.0223) & 0.853 (0.0323) & 0.823 (0.0228) & 0.96 (0.0079) & 0.944 (0.0093) & 0.922 (0.019) & 0.875 (0.02) & 0.832 (0.0294) & 0.782 (0.0247) \\ 
   & & MIDA & 0.959 (0.0128) & 0.941 (0.0148) & 0.925 (0.0193) & 0.883 (0.0158) & 0.827 (0.0123) & 0.801 (0.0257) & 0.964 (0.0124) & 0.948 (0.0163) & 0.925 (0.0179) & 0.873 (0.0225) & 0.831 (0.0353) & 0.761 (0.0277) \\ 
   & & GAIN & 0.961 (0.0113) & 0.947 (0.0162) & 0.927 (0.0221) & 0.889 (0.0234) & 0.851 (0.0196) & 0.807 (0.0245) & 0.969 (0.0134) & 0.946 (0.0137) & 0.924 (0.0198) & 0.882 (0.016) & 0.817 (0.0249) & 0.767 (0.0185) \\ 
   & & DPNB & 0.968 (0.0117) & 0.961 (0.0157) & 0.952 (0.0111) & 0.932 (0.0186) & 0.91 (0.0219) & 0.875 (0.0202)  & 0.973 (0.0107) & 0.964 (0.0125) & 0.95 (0.0181) & 0.939 (0.016) & 0.897 (0.0202) & 0.862 (0.0298) \\ 
%   & & DPNB-T & 0.962 (0.0097) & 0.948 (0.0111) & 0.933 (0.0191) & 0.911 (0.0154) & 0.879 (0.0214) & 0.83 (0.0151) & 0.967 (0.0122) & 0.961 (0.0117) & 0.935 (0.0158) & 0.906 (0.0095) & 0.877 (0.0192) & 0.824 (0.0314) \\ 

		\midrule
\multirow{7}{*}{Wine Quality} &  \multirow{7}{*}{F1 Score} & RF & 0.587 (0.0087) & 0.544 (0.0133) & 0.525 (0.0149) & 0.498 (0.018) & 0.448 (0.0167) & 0.401 (0.0192) & 0.594 (0.0118) & 0.544 (0.0122) & 0.528 (0.0114) & 0.496 (0.014) & 0.445 (0.0133) &  0.401 (0.0148) \\ 
  & & MICE & 0.587 (0.0145) & 0.548 (0.0099) & 0.515 (0.0163) & 0.481 (0.0263) & 0.438 (0.0271) & 0.403 (0.0137) & 0.592 (0.0108) & 0.548 (0.0091) & 0.529 (0.0108) & 0.486 (0.02) & 0.437 (0.0219) & 0.404 (0.0284) \\ 
  & & KNN & 0.592 (0.0115) & 0.55 (0.0097) & 0.52 (0.0114) & 0.493 (0.0117) & 0.466 (0.0238) & 0.419 (0.0146) & 0.594 (0.0105) & 0.555 (0.0106) & 0.539 (0.017) & 0.499 (0.0139) & 0.457 (0.0245) & 0.427 (0.0176) \\ 
  & & MIDA & 0.589 (0.011) & 0.548 (0.0159) & 0.526 (0.0147) & 0.494 (0.0259) & 0.459 (0.031) & 0.423 (0.0186) & 0.592 (0.0102) & 0.552 (0.0088) & 0.525 (0.0143) & 0.5 (0.0164) & 0.456 (0.0221) & 0.421 (0.0261) \\ 
  & & GAIN & 0.589 (0.0106) & 0.551 (0.0124) & 0.515 (0.0183) & 0.489 (0.0198) & 0.448 (0.0273) & 0.407 (0.0161) & 0.586 (0.012) & 0.55 (0.0097) & 0.526 (0.0166) & 0.493 (0.025) & 0.453 (0.0243) & 0.409 (0.0152) \\ 
  & & DPNB & 0.597 (0.0092) & 0.566 (0.0077) & 0.547 (0.014) & 0.528 (0.0149) & 0.51 (0.0082) & 0.488 (0.0078)  & 0.596 (0.0108) & 0.567 (0.0111) & 0.544 (0.0104) & 0.527 (0.0135) & 0.511 (0.013) & 0.487 (0.0108)  \\ 
%   & & DPNB-T & 0.587 (0.0099) & 0.563 (0.006) & 0.542 (0.0134) & 0.527 (0.015) & 0.508 (0.0084) & 0.486 (0.0085) &  0.585 (0.0114) & 0.555 (0.0066) & 0.537 (0.0117) & 0.525 (0.0137) & 0.502 (0.0101) & 0.479 (0.0109) \\ 

 \bottomrule
	\end{tabular}}
  \caption{Average of 10 times repeated 10-folds cross-validated metrics for four UCI datasets with different missingness patterns and missing rates. Estimated standard errors of the average cross-validated scores are shown in parentheses.}
  \label{tab:cl_result}
\end{table}

\begin{table}[t!]
\centering
\footnotesize{
\begin{tabular}{cccccc}
  \toprule
 & 50\% & 60\% & 70\% & 80\% & 90\% \\ 
  \midrule
RF+L & 0.161 (0.032) & 0.168 (0.0306) & 0.168 (0.0322) & 0.182 (0.0344) & 0.177 (0.0467) \\ 
  RF+R & 0.112 (0.0632) & 0.123 (0.0585) & 0.124 (0.0647) & 0.118 (0.0655) & 0.111 (0.0789) \\ 
  MICE+L & 0.146 (0.0346) & 0.151 (0.0322) & 0.157 (0.0365) & 0.168 (0.0409) & 0.171 (0.0595) \\ 
  MICE+R & 0.006 (0.0118) & 0.005 (0.0117) & 0.007 (0.0163) & 0.008 (0.0186) & 0.005 (0.0194) \\ 
  KNN+L & 0.141 (0.0234) & 0.146 (0.0324) & 0.149 (0.0362) & 0.161 (0.0446) & 0.145 (0.0587) \\ 
  KNN+R & 0.118 (0.0492) & 0.121 (0.0528) & 0.127 (0.0527) & 0.131 (0.0531) & 0.131 (0.0618) \\ 
  MIDA+L & 0.11 (0.0306) & 0.11 (0.0338) & 0.122 (0.0403) & 0.14 (0.0489) & 0.13 (0.0641) \\ 
  MIDA+R & 0.166 (0.0256) & 0.167 (0.0324) & 0.178 (0.0215) & 0.178 (0.0232) & 0.174 (0.0334) \\ 
  GAIN+L & 0.15 (0.0316) & 0.158 (0.0316) & 0.171 (0.0289) & 0.181 (0.0343) & 0.171 (0.0602) \\ 
  GAIN+R & 0.16 (0.0387) & 0.172 (0.0216) & 0.168 (0.0347) & 0.174 (0.0291) & 0.174 (0.0353) \\ 
  DPNB  & 0.169 (0.0223) & 0.171 (0.0182) & 0.171 (0.0248) & 0.179 (0.0309) & 0.177 (0.042) \\
   \bottomrule
\end{tabular}
\caption{Average of F1-scores over 100 replications for the semiconductor dataset with different ratios between training and test sets. Estimated standard errors of F1-scores are shown in parentheses.}
\label{tab:sssc_cls_f1score}
}
\end{table}

\end{appendices}
% \section{Derivation of variational lower bound for DPMM with missing data}
\end{document}